\documentstyle[psfig,prb,aps]{revtex}
\begin{document}
\draft

\twocolumn[\hsize\textwidth\columnwidth\hsize\csname
@twocolumnfalse\endcsname

\title{
Orbital-free molecular dynamics simulations of melting in Na$_8$ and Na$_{20}$:
melting in steps.
}
\author{Andr\'es Aguado\footnote{Electronic mail: aguado@jmlopez.fam.cie.uva.es}, 
Jos\'e M. L\'opez and Julio A. Alonso}
\address{Departamento de F\'\i sica Te\'orica,
Universidad de Valladolid, Valladolid 47011, Spain}
\author{Malcolm J. Stott}
\address{Department of Physics, Queen's University, Kingston,
Ontario K7L 3N6, Canada}
\maketitle
\begin{abstract}
The melting-like transitions of Na$_8$ and Na$_{20}$ are investigated by
{\em ab initio} constant energy molecular dynamics simulations,
using a variant of the Car-Parrinello method which employs an explicit
electronic kinetic energy functional of the density, thus avoiding the use of
one-particle orbitals. Several melting indicators are evaluated in order to
determine the nature of the various transitions, and compared with other
simulations.  Both Na$_8$ and Na$_{20}$ melt over a wide temperature
range. For Na$_8$, a transition is observed to begin at $\sim$ 110 K, between a rigid phase and a
phase involving isomerizations between the different permutational isomers
of the ground state structure. The ``liquid'' phase is completely
established at $\sim$ 220 K. For Na$_{20}$, two transitions
are observed: the first, at $\sim$ 110 K, is associated
with isomerization transitions between those permutational isomers of the
ground state structure which are obtained by interchanging the positions
of the surface-like atoms; the second, at $\sim$ 160 K, involves a structural
transition from the ground state isomer to a new set of isomers with the
surface molten. The cluster is completely ``liquid'' at $\sim$ 220 K.

\end{abstract}
\pacs{PACS numbers: 36.40.Ei 64.70.Dv}

\vskip2pc]

\section{Introduction}

Recent experimental advances have made possible the study of the melting
phenomenon in metal clusters. Martin \cite{Mar96} measured the melting
temperatures T$_m$ of large sodium clusters by observing the disappearance of
the atomic shell structure with increasing temperature, and determined 
the dependence of T$_m$ on cluster size. More recently, Haberland and
coworkers \cite{Sch97} determined the heat capacity and melting
temperature of free sodium clusters containing from 50 to 200 atoms 
by studying the temperature dependence of the photofragmentation mass
spectrum. Electron diffraction patterns of trapped clusters \cite{Joe98} may be
useful for distinguishing different stages in the melting process.

For many years, computer simulations have been the main guide to an 
understanding of melting-like transitions in small finite systems.
Molecular dynamics (MD) calculations based on phenomenological potentials have
been used to study the solid-like to liquid-like phase transitions in rare gas,
\cite{Jel86} alkali halide, \cite{Ros93,Cal98}
and metal \cite{Bul92,Guv93,Rey93,Pot94,Fos95,Bul97,Cle98,Cal99}
clusters. The unification of density functional theory (DFT) and molecular dynamics
formulated by Car and Parrinello \cite{Car85} allows for an
explicit treatment of the electronic degrees of freedom. Such {\em ab initio}
molecular dynamics calculations have been performed by R\"othlisberger
and Andreoni \cite{Rot91} for Na$_n$ (n=2--20) microclusters at several 
temperatures, and very recently, Rytk\"onen {\em et al.} \cite{Ryt98} have
presented results for the melting of Na$_{40}$. These studies
make use of the Kohn-Sham (KS) version of DFT, \cite{Koh65} but large
computational savings can be obtained if an orbital-free method, based solely
in the electronic density $n(\vec r)$, is used instead. Madden and coworkers
\cite{Pea93} have demonstrated the value of orbital-free 
methods for the study of bulk metallic systems. Shah {\em et al.} \cite{Sha94}
and Govind {\em et al.} \cite{Gov95} have applied the  approach to
study the structures of small metal and covalent clusters.
Blaise {\em et al.} \cite{Bla97} have performed orbital-free 
calculations of some static and dynamic properties of sodium clusters with
sizes up to 274 atoms, and Vichare and Kanhere \cite{Vic98} have studied the
melting of Al$_{13}$. 
In this paper  we describe the study of solid-like to liquid-like phase
transitions of Na$_8$ and Na$_{20}$ microclusters using constant energy
orbital free molecular dynamics simulations. 

In the next section we present the theoretical details of the method. 
The results are presented and discussed in section III. Section IV summarizes
the main conclusions from this study.

\section{Theory}

\subsection{Car-Parrinello Molecular Dynamics with Orbital-Free Energy Density
Functionals}

The orbital-free molecular dynamics method is a Car-Parrinello total
energy scheme which uses an explicit
kinetic-energy functional of the electron density, and has the valence electron
density as the dynamic variable.
Now we describe the main features of the energy functional.
The orbital-free calculational scheme has been described at length in Refs. 
\onlinecite{Pea93,Sha94,Gov95,Bla97}, 
so we will give just a brief description here.

The ground state energy is a functional of the valence electron density
$n(\vec r)$, and a function of the ion positions $\vec R_n$
($n = 1, 2, \dots N$), with the following form (Hartree atomic units will be
used through the paper):
\begin{eqnarray}
E[n,\vec R] = T_s[n] + \frac{1}{2} \int \int \frac{n(\vec r)n(\vec r^ {'})}{\mid
\vec r - \vec r^ {'} \mid}d\vec r d\vec r^ {'} \nonumber \\
+ \int n(\vec r)V_{ext}(\vec r)d\vec r + E_{XC}[n] + E_{ion-ion}[\vec R],
\end{eqnarray}
where the different terms are: the kinetic energy functional for a noninteracting
inhomogeneous electron gas, the classical electron-electron interaction
energy, the interaction
energy between the valence electrons and the external potential
provided by the instantaneous ionic configuration, the exchange-correlation
energy functional, and the classical Coulomb repulsion between positive ions.
Three key approximations in the energy functional involve T[n],
E$_{XC}$[n], and the electron-ion interaction.
The electronic kinetic energy functional used in this work 
corresponds to the gradient expansion
around the homogeneous limit through second order, \cite{Mar83}
\begin{equation}
T_s[n] =
T^{TF}[n] + \lambda T^W[n],
\end{equation}
where the first term is the Thomas-Fermi
approximation,
\begin{equation}
T^{TF}[n] = \frac{3}{10}(3\pi^2)^{2/3}\int_{\Omega}n(\vec r)^{5/3}d\vec r,
\end{equation}
and T$^W$ (the Weizs\"acker term) is the lowest-order gradient correction to
T$^{TF}$
\begin{equation}
T^{W}[n] = \frac{1}{8}\int_{\Omega}\frac{\mid \nabla n(\vec r) \mid^2}{n(\vec r)}d\vec r,
\end{equation}
taking some account of inhomogeneities in the electron density.
Different values have been proposed in the literature for the constant $\lambda$.
The value adopted here, $\lambda = \frac{1}{9}$, corresponds to the 
limit of a slowly varying $n(\vec r)$
and has a number of desirable properties. \cite{Hoh64,Yan86,Per92}
The local density approximation (LDA) is used
for the exchange-correlation functional.
We use the Perdew and
Zunger parametrisation \cite{Per81} of the electron gas results of Ceperley
and Alder. \cite{Cep80}
The external field contains the electron-ion interaction,
$V_{ext}(\vec r) = \sum_i v(\vec r -\vec R_i)$, where we take for $v$
the local pseudopotential of Fiolhais {\em et al.}, \cite{Fio95} which
reproduces very well the properties of bulk sodium and has shown good
transferability to sodium clusters. \cite{Nog96}
Although some progress has been made recently for including the effects
of nonlocality in the pseudopotential in orbital free schemes,
\cite{Wat98} these effects are expected to be small for sodium. 
The ion-ion interactions are treated using the usual Ewald method. \cite{Ewa21}

For a given set of ion positions \{$\vec R$\}, that is for a given external potential
$V_{ext}$, the ground state is obtained from the variational principle:
\begin{equation}
\frac{\delta}{\delta n(\vec r)} (E[n,\vec R] - \mu \int n(\vec r)d\vec r ) = 0,
\end{equation}
where $\mu$ is the electron chemical potential chosen to give the desired number
of electrons N$_e$. 
We have found it convenient to work in terms of a single effective
orbital $\psi(\vec r)$ rather than $n(\vec r)$, where
\begin{equation}
n(\vec r) = \mid \psi(\vec r) \mid^2,
\end{equation}
and to vary $\psi$ rather than $n$. This
has the advantage of maintaining $n$ nonnegative if $\psi$ is real.
The cluster of interest is placed in a unit cell of a superlattice of
volume $\Omega$, and the set of plane waves periodic in the superlattice
is used to expand $\psi$. The expansion coefficients C$_{\vec G}$ (where
$\vec G$ is a reciprocal lattice vector of the superlattice) are considered
as generalized coordinates of a set of fictitious classical particles each of
mass $m$.\cite{Car85} The Lagrangian for the whole system of electrons and
ions is
%
\begin{equation}
L(C,\dot{C},\vec R) = \sum_{\vec G} \frac{1}{2}m\mid \dot{C}_{\vec G} \mid^2
+\sum_{i=1}^N \frac{1}{2} M_i \dot{\vec{R_i}}^2 - E[n,\vec R],
\end{equation}
and the constraint to be considered is
\begin{equation}
\int_{\Omega}d\vec r n(\vec r) = \Omega\sum_{\vec G}C_{\vec G}^*C_{\vec G} = 
N_e,
\end{equation}
where $N_e$ is the number of electrons per supercell, and $M_i$
is the mass of the ith ion. 

The explicit functional of the density we have used for the electronic
energy is much superior in computational speed and memory
requirements to the conventional KS orbital approach,
allowing the treatment of larger systems for longer simulation times.
Those computational savings can also be used to perform simulations for a
larger number of different temperatures, allowing better 
identification of transition temperatures.
The present restriction to local pseudopotentials being not a serious
matter for sodium, the most important difference between the
orbital-free and KS approaches is the treatment of the independent
particle kinetic energy. T$_s$[n] is computed exactly in the KS
approach,
whereas the orbital-free approach
draws on approximate density functionals based on a few known limiting
cases. Our choice for T$_s$[n] with $\lambda = \frac{1}{9}$ is
appropriate for a slowly varying density, whereas $T_s [n] = T^W [n]$ is
believed to give the limit of rapidly varying density. Linear
response theory gives T$_s$[n] when the variations in electron density
about the mean density are small, and some functionals attempt to
combine these different limits. \cite{Pea93} Our functional would seem
to be appropriate for the smooth pseudoelectron density of a Na cluster,
but since little is known about the range of validity of the various
functionals, we have performed test calculations on some simple Na
systems and compared the results with those of other approaches.

Table I presents results for equilibrium bond lengths and binding
energies per atom for the Na dimer and an octahedral Na$_6$ cluster
calculated within three different approximations.
All the calculations use the LDA for
exchange and correlation, but differ in their treatment of the electron-ion
interaction and the electron kinetic energy. The results \cite{Nog96} of KS
all-electron calculations are given along with those of KS
calculations which use our choice of local pseudopotential, and comparison
of these results tests the pseudopotential of Fiolhais. \cite{Fio95}
We see that the use of the pseudopotential overestimates the bond lengths by
2--3 percent, but the increased
bond length in Na$_6$ over that in the dimer is reproduced. The binding
energies, with respect to spin-polarized free atoms, are also in
reasonable agreement. The third set of calculations uses the same
pseudopotential and the orbital-free kinetic energy functional.
Comparison of the second and third sets of results tests the
orbital-free T$_s$[n]. The approximate T$_s$[n]
overestimates the bond length of the dimer by 4 \% but the results for
Na$_6$ are in very good agreement. The binding energies (with respect to
spin-polarized free atoms) are
overestimated but the increased binding in Na$_6$ is reproduced.
The agreement between orbital-free and KS calculations for
the ground state structure and the energetic ordering of the low lying
isomers of Na$_8$ which is detailed in section III, and the
reasonably accurate interatomic distances we obtain, provide
additional support to our approach for studying the melting-like
transitions of Na clusters.  This agreement suggests that the
orbital-free approach
gives a potential energy surface with the correct features for
interatomic separations near the equilibrium distances, and this is enough 
to determine the details of the melting transition. The
discrepancies in the binding energies are less important for the
dynamics provided we do not consider evaporation events.

The main approximation of an orbital-free method is the neglect of quantum
shell effects. Specifically, the method gives energies that vary smoothly
as a function of cluster size, without showing the oscillations associated
with electronic shell closures. On the other hand ionic shell effects due to the geometrical
arrangement of ions are present. The experimental results of
Haberland {\em et al.} \cite{Sch97} seem to indicate that both effects are
relevant to the melting of sodium clusters, although their relative importance
is not yet known. Nevertheless, the good performance
of the orbital-free method mentioned above, together with the good agreement
between the melting dynamics of Na$_8$ and Na$_{20}$ predicted by our model
and that obtained from the KS calculations of R\"othlisberger 
and Andreoni \cite{Rot91} (see section III), support the orbital-free method as
a valuable starting point. Extended Thomas-Fermi models usually give equilibrium
structures that are quite spherical, and do not have the deformations found in
small clusters of size intermediate between two electronic shell closures. By restricting
our study to closed-shell clusters, which are known to be quite spherical,
we minimize the probability of finding wrong zero-temperature isomers.

The calculations for Na$_8$ and Na$_{20}$ have been performed using a
cubic supercell of edge 63.58 a.u. with a 64 $\times$ 64 $\times$ 64 mesh 
and a plane wave energy cut-off for $\psi$ of 10 Rydbergs. 
The test calculations performed for Na$_2$ and Na$_6$ showed that with this
cutoff, equilibrium bond lengths are converged within 0.01 \AA \
and binding energies within
0.002 eV, which we consider sufficient for our purposes.
The equations of motion were integrated using the Verlet
algorithm \cite{Ver65} for both the electrons and the ions with a time
step ranging from $\Delta$t = 1 $\times$ 10$^{-15}$ sec. for the
simulations performed at the lowest temperatures, to $\Delta$t = 0.85
$\times$ 10$^{-15}$ sec. for those performed at the highest ones. The
fictitious electron mass ranged from 1.15 $\times$ 10$^8$ a.u. for the
shorter time steps, to 1.75 $\times$ 10$^8$ a.u. for the longest.
These choices resulted in a conservation
of the total energy better than 0.1 \%. The first step in the simulation
was the determination of the low temperature ground state structures
(GS) of Na$_{8}$ and Na$_{20}$ using the dynamical simulated annealing 
technique \cite{Car85} by heating the clusters to 600 K and then slowly cooling
them.  Next, several molecular dynamics simulation
runs at different energies were performed in order to obtain the caloric curve.
The initial relative positions of the atoms in the cluster
for the first run were taken by slightly deforming the equilibrium geometry.
The final configuration for each run served as the starting geometry for the
next run at a different energy. The initial velocities for every new run were
obtained by scaling the final velocities of the preceding run.
The total simulation times were 8 ps for the lowest temperatures for which the
clusters are very rigid (T $<$ 60 K), and 20--22 ps for
temperatures larger than 60 K. Longer runs of 50--60 ps were performed for
specific temperatures.  The first 2 ps of each run were not included in the
various time averages. After the velocity scaling performed at the beginning
of each new trajectory, the internal cluster temperature (see
next section) was always an oscillating function of time with a decreasing
envelope during approximately the first 2 ps of the simulation, after which  
equilibration is achieved. The reason for such a short equilibration time seems
to be that the velocity scaling never increased the cluster temperature in
more than 20 K, at least in the transition region. Equilibration was also
checked by comparing the caloric curve and the specific heat curve (see
following section). The locations of the specific heat peaks were in coincidence
with the slope changes in the caloric curve, which is a sign of correct
equilibration.

\subsection{Analysis of the Molecular Dynamics}

In order to characterize the thermal behavior of the clusters as a function of
increasing internal energy, and the solid-like to liquid-like transitions, we
monitor: (a) global quantities that are calculated from time
averages over a whole trajectory at a given energy; (b) time dependent
quantities  that are calculated from averages over well separated time origins
along a single trajectory.

First, we define the ``internal temperature'' T of the cluster as \cite{Jel86,Sug91}
\begin{equation}
T = \frac{2}{(3N-6)k_B}<E_{kin}>_t,
\end{equation} 
where $k_B$ is the Boltzman constant, $E_{kin}$ is the ionic
kinetic energy, and $< >_t$ represents the time average over a whole trajectory.
All the global quantities described below will be plotted as functions of this
internal temperature.

The degree of mobility of atoms - a sort of index of rigidity of the 
cluster - can be characterized by the relative root-mean-square (rms) bond
length fluctuation $\delta$, defined by \cite{Jel86,Sug91}
\begin{equation}
\delta = 
\frac{2}{N(N-1)} \sum_{i>j} \frac{(<R_{ij}^2>_t - <R_{ij}>_t^2)^{1/2}}{<R_{ij}>_t},
\end{equation} 
where $R_{ij} = \mid \vec R_i - \vec R_j \mid$. This quantity changes 
abruptly at isomerization or melting transitions, and for the bulk, a sharp
increase in $\delta$ gives the Lindemann criterion for melting.

As another indicator of the melting-like transition we calculate the specific
heat defined by \cite{Jel86,Sug91}
\begin{equation}
C = [N - N(1-\frac{2}{3N-6})<E_{kin}>_t ~ <E_{kin}^{-1}>_t]^{-1}.
\end{equation} 
This magnitude is related to the fluctuations in the ionic kinetic energy, and
has peaks (corresponding to slope changes in the caloric curve)
associated to some phase transitions.

The mean square displacement, R(t), defined by
\begin{equation}
R(t) = \frac{1}{Nn_t}\sum_{j=1}^{n_t}\sum_{i=1}^N [\vec r_i(t_{0_j} + t) -
\vec r_i(t_{0_j})]^2, 
\end{equation} 
where $n_t$ is the number of time origins ($t_{0_j}$) considered along a trajectory,
is a time-dependent quantity that also serves as a measure of the rigidity of
the cluster. \cite{Jel86}
The slope of R(t) for large t is proportional to the
diffusion coefficient. Thus, a flat R(t) curve is indicative of a
solid-like cluster with the constituent atoms vibrating about their
equilibrium positions; when diffusive motion of the atoms starts,
the slope of R(t) becomes positive. We shall see in the next section
that it is useful to separate the atoms of Na$_{20}$ into two subsets of
N$_c$=2 ``core'' and N$_s$=18 ``surface'' atoms.  We can then define
partial R(t)'s, restricted to core (R$_c$(t)) and surface (R$_s$(t)) atoms, 
respectively, and the differences between these will be indicators of
different degrees of rigidity in the two groups of atoms for a given
internal cluster temperature.

Recently, \cite{Bon97} the ``atomic equivalence indexes'' 
\begin{equation}
\sigma_i(t) = \sum_j \mid \vec r_i(t) - \vec r_j(t) \mid, 
\end{equation} 
which contain very detailed
structural information, have been introduced. The degeneracies in $\sigma_i(t)$
are due to the specific symmetry of the isomer under consideration, and the
variations in the time evolution of $\sigma_i(t)$ allow for a detailed
examination of the melting mechanism. 

\section{Results and Discussion}

\subsection{Na$_8$}

The calculated GS structure of Na$_8$
is shown in Fig. 1(a). It is a dodecahedron (D$_{2d}$ symmetry), a result which
is in agreement with the KS-LDA calculations. \cite{Rot91,Mar85,Pac97} The
stellated tetrahedron (T$_d$ symmetry) is the GS structure in 
all-electron SCF-CI calculations, \cite{Bon88,Spi89} although the energy
difference between the isomers is very small indeed. It is noteworthy that
the orbital-free-LDA calculations lead to the same
energetic ordering of isomers (and also to similar energy differences between
isomers) as the KS-LDA.

Figure 2 shows the results for the calculated global quantities C
(fig. 2(a)) and $\delta$ (fig. 2(b)) as functions of internal
cluster temperature, and the time evolution of the mean square
displacement, R(t), for three representative temperatures (fig. 2(c)).
The specific heat shows a peak centered at a temperature T$_{m_1} \approx $
110 K, which correlates with the temperature region in which
$\delta$ experiences an almost stepwise increase, and marks the onset of the
melting transition.
This is followed by a steady increase of $\delta$(T), until a levelling
off occurs at T$_{m_2} \approx$ 220 K,  
indicating that the liquid state is fully developed. The melting
transition of Na$_8$ from a rigid form, in which the atoms
merely oscillate about the equilibrium configuration, to a
``fluid'' form characterized by uncorrelated motion of the atoms, is
spread over a range of temperatures T=T$_{m_1}$-T$_{m_2}$. Fig. 2(c)
shows that for a temperature T=34 K, R(t) has zero slope at long times,
reflecting the oscillatory motion of the atoms. At T=111 K the diffusive motion
of the atoms in the cluster begins, resulting in a positive slope which
increases with temperature.

The quantities shown in Fig. 2 indicate that Na$_8$ undergoes a melting-like
transition in the temperature range 110-220 K.
In order to get a better understanding
of the melting, the short-time averages of the
``atomic equivalence indexes'', $<\sigma_i(t)>_{sta}$,
have been calculated and the cluster evolution during the
trajectories has been followed visually using computer graphics.
The $<\sigma_i(t)>_{sta}$ curves are presented in Fig. 3 for three
representative values of T. At the lowest temperature T=34 K (fig. 3(a)) 
there are two distinct groups of nearly equivalent atoms in the D$_{2d}$
GS structure: four atoms with coordination 4, and four with coordination
5. The curve crossings within a group of nearly equivalent atoms display
oscillations of the atoms around their equilibrium positions, but
there are no crossings of curves from different groups. At a higher
temperature T=78 K (not shown), but still lower than T$_{m_1}$, the
oscillations have larger amplitudes, but the two types of atoms do not
yet mix. At T=111 K, which marks the beginning of the step in $\delta$,
interchanges between atoms with different coordination begin, but are
still scarce. The dynamics over 60 ps (fig. 3b) illustrates 
that the atoms in the cluster can still be separated into two sets with 
4 and 5 coordination during most of the simulation time, there being 
just four interchange events.  The movies show
that the onset of melting in Na$_8$ is associated with isomerization
transitions between the different permutational isomers of the D$_{2d}$
GS structure, which come with significant distortions of
the cluster from the GS geometry. The transitions become more frequent
and the distortions increasingly severe with increasing temperature 
giving rise to the steady increase observed in $\delta$, until at
T$ \approx$ 220 K (fig. 3c)
and higher temperatures the ground state isomer is seen only occasionally
in the movies. Snapshots of the trajectories at these temperatures show 
that the cluster sometimes adopts elongated configurations with Na or
Na$_2$ subunits joined to the cluster by a single bond. 
This is reflected in figure 3c by the larger values of $<\sigma_i>_{sta}$. All
the atoms diffuse across the cluster volume so that the different
$<\sigma_i>_{sta}$ curves are all mixed and the ``liquid''
phase is fully established. While we have
not observed evaporation of fragments at this temperature (at least within
20 picoseconds), the trajectories suggest that Na or Na$_2$ subunits may
evaporate at higher temperatures from the ends of a ``cigar-like'' cluster.
However, our energy functional may not be reliable for such events.

Detailed simulations of the melting-like transition of Na$_8$ have been
performed by Bulgac and coworkers \cite{Bul92} using constant-temperature
MD simulations with a phenomenological interatomic
potential, and by Poteau, Spiegelmann and Labastie, \cite{Pot94}
using Monte Carlo (MC) simulations with a tight-binding Hamiltonian to describe
the electronic system; recently, Calvo and Spiegelmann \cite{Cal99}
(CS) have performed also
MC simulations employing an empirical potential.
Aside from the differences between MC and MD methods,
the main difference between those calculations and our work is the
interatomic potential. That employed by Bulgac has free parameters
fitted to properties
of bulk sodium, and while the tight-binding calculations of PSL deal
explicitly with the electronic system, some overlap integrals were fitted
to {\em ab initio} results for the sodium dimer and tetramer.  Since we
consider the electronic degrees of freedom explicitly and find the forces
on ions through the Hellman-Feynman theorem, the present calculations
should be considered an advance over empirical or semiempirical
methods. Bulgac predicts a transition temperature of about
100 K from a
solid-like to a glassy phase of the cluster in which the atoms stay
close to their equilibrium positions for relatively long periods with 
occasional atom interchanges. The tight binding treatment \cite{Pot94}
gives a melting-like transition at $\sim$ 200 K, whereas CS give an approximate
value of 80--100 K for T$_m$, more similar to the values of Bulgac and 
ourselves. Bulgac also predicts a
boiling transition at 900 K but we do not expect our approach to be
reliable at such high temperatures where evaporation is a factor. However,
the ``solid-to-glassy'' transition is similar to the behaviour we have
found at the onset of melting at $\sim$ 110--130 K, where the atoms
remain in their equilibrium positions for relatively long periods, 
only swapping positions from time to time with a frequency which increases
with temperature. For temperatures above 220 K where the value of $\delta$
increases smoothly with T, we find that the atoms are very mobile and
the cluster can be considered liquid.

R\"othlisberger and Andreoni (RA) \cite{Rot91} have performed finite
temperature simulations for Na$_8$ using the {\em ab initio} Car-Parrinello
MD method with pseudopotentials for the electron-ion interaction and the
LDA for the exchange-correlation energy functional. Thus, the main
difference from the present calculations is our use of an approximate
electronic kinetic-energy functional. They deduced from a simulation at
T=240K a value of $\delta$ still less than the 0.1 value which is taken
to indicate the onset of melting according to the Lindeman criterion.
The full KS method gives a more accurate description of the electronic structure 
than our orbital-free approach, but we feel that the apparently
contradictory results for $\delta$ can be reconciled. The total simulation
time of RA for Na$_8$ was t=6 ps, corresponding to about $\sim$ 10
vibrational periods for sodium. \cite{Mar96} Our calculated value of
$\delta$ at T=230 K, if we average over the first 6 ps of the simulation
is 0.098, in agreement with the full KS results.  Due to the nature of
the melting transition in Na$_8$, involving a repeated swapping of the
atom positions, 6 ps may not be long enough to obtain a converged value
for $\delta$. In fact, we chose the total simulation time of t=20 ps after
recognizing that preliminary simulations performed with t=8 ps did not
yield a converged result for $\delta$. With t=20 ps, further increases
only lead to minor changes in $\delta$, at least when the cluster is clearly
liquid-like. Nevertheless, we can not guarantee that fully converged values
of $\delta$ have been obtained in the transition region.

\subsection{Na$_{20}$}

We find the ground state structure of Na$_{20}$ to be a single-capped
double icosahedron (fig. 1(b)), again in good agreement with the KS-LDA
calculations of RA, \cite{Rot91} who found this structure as an isomer
almost degenerate with their GS structure.
The variation of the specific heat with temperature, which is shown
in fig. 4(a), displays two maxima around 110 K and 170 K.
The relative rms bond length fluctuation is given in fig. 4(b).
For small temperatures the curve $\delta (T)$ has a small positive slope
reflecting the thermal expansion of the solid-like cluster, but at higher
temperatures there are two abrupt increases at $T_{m_1} \approx
110 K$ and $T_{m_2} \approx 160 K$, and a levelling off at T$_{m_3} \approx
220 K$, indicating
that the melting of Na$_{20}$ occurs in several stages over the
range of temperatures T$_{m_1}$--T$_{m_3}$. The two peaks in the specific
heat are in rough correspondence with the two steps in $\delta$(T), but
the levelling off in $\delta$ does not seem to be associated with any
pronounced feature in the specific heat. The mean square displacement for
the particles in the cluster is plotted as a function of time in Fig.
4(c) for four different temperatures.  The curve corresponding to T=55 K
shows a clear levelling off after a small initial rise, reflecting the
solid-like nature of the cluster at that temperature. For T = 109 K and
higher temperatures, $R(t)$ shows a linear increase implying a finite
diffusion coefficient and the emergence of liquid-like properties.

In order to investigate the nature of the transitions we divide the 20
atoms of the cluster into two subsets: the two internal ``core'' atoms
of the ground state structure, and the 18 peripheral
``surface'' atoms. The partial R$_c$(t) and R$_s$(t) for the temperature
at which the first step in $\delta$ begins are shown in fig. 5.
The R$_s$(t) curve shows the typical diffusive behavior of liquid-like
phases, whereas R$_c$(t), after the initial rise, oscillates with an
average slope near zero. Differences in the mobility of the surface and
core atoms are clearly indicated, the surface atoms undergoing diffusive
motion at a lower temperature (T$_{m_1} \approx 110$ K) than the core
atoms. Snapshots at regular time intervals for the runs between
T=T$_{m_1}$ and T=T$_{m_2}$ show the two central atoms
oscillating around their initial positions, while surface atoms interchange
their positions in the cluster without destroying the double-icosahedral
symmetry. The picture is similar to the onset of melting in Na$_8$.
The swapping of surface atom positions causes significant
distortions of the double-icosahedron geometry, but after each interchange
the cluster returns to the ground state structure.
Therefore, we identify the transition at T$_{m_1} \approx 110$ K as
an isomerization-like transition in which the cluster begins to visit those
permutational isomers of the ground state which involve the
interchange of positions between surface atoms. In our view, 
this fluctuating phase of Na$_{20}$ can not be associated with surface
melting, because the double-icosahedral symmetry persists and the surface
disorder is temporary. 

The next stage is seen in snapshots from the runs performed at
temperatures between T$_{m_2}$ and T$_{m_3}$. The second
transition involves the motion of one of the central atoms out to a peripheral
position, the double-icosahedral symmetry is lost, and most of the snapshots
show a structure with a single central atom (see an example in fig. 1c).
The central core atom exchanges with one of the 19
surface atoms at a slower rate than the interchanges
between surface atoms, and so converged values of $\delta$ require
longer simulations (50--60 ps).  The 19 surface
atoms are very mobile and the structure of the cluster fluctuates a great
deal. The exploration at these temperatures of a new structure
leads to the second peak in the specific heat, and we identify T$_{m_2}
\approx 160 K$ with an isomerization-like transition in which new
(19+1) structures are explored. As the temperature is increased 
from T$_{m_2}$ to T$_{m_3}$, the interchanges of a core atom and a surface
atom become more frequent, leading to the steady increase in $\delta$,
and a cluster surface which has melted in the sense that the structure is
now very fluid, and the surface disorder is large.

In fig. 6 we show the short-time averaged $\sigma_i(t)$ for several
representative temperatures. At T=32 K (fig. 6(a)) the cluster is
solid-like and $<\sigma_i(t)>_{sta}$ serves to identify groups of ``nearly''
equivalent atoms.  Starting from the bottom, these are: the two central
atoms, the five atoms of the central pentagon, the 10 atoms in the upper
and lower pentagons, and finally the two axial atoms and the capping
atom. At T=72 K (not shown)
the different groups can still be recognized, but fig. 6(b)
shows that at 137 K there is a group of two central atoms, but 
the rest are no longer distinguishable. At T=163 K (fig. 6(c))
we see that the cluster has a single central atom for most of the
simulation time (56 ps), the 19 atom surface is melted, and occasionally an
interchange between the central atom and one of the surface atoms occurs.
At higher temperatures the interchange rate increases.

Simulations of the melting-like transitions of Na$_{20}$ have also been performed
by Bulgac \cite{Bul92}, PSL \cite{Pot94} and CS.\cite{Cal99} As for Na$_8$, 
the ground state structure of Na$_{20}$ predicted by PSL
is different from that of KS-LDA calculations, although it may also be divided
in ``core'' and ``surface'' atoms. They also predict surface melting at
a temperature of T$_s$=190 K, and a melting temperature of T$_m$=300 K.
The discrepancies between these quantities and the corresponding
quantities we obtain should again be attributed to the differences in the
interatomic interactions. Once more our detailed description of  
the onset of the melting transition and the full establishment
of a liquid-like phase are in better agreement with the results of Ju and
Bulgac. \cite{Bul92} CS predict the same ground state structure found in this
work for Na$_{20}$. Their calculations also predict a two-step melting
mechanism, with T$_{m_1}\approx$ 100 K and T$_{m_2}\approx$ 200 K.\cite{Cal99}
RA \cite{Rot91} have
performed two short {\em ab initio} KS-LDA molecular dynamics simulations
of Na$_{20}$. At T=350 K, the lowest temperature they considered, they found
that the mobility of the atoms in Na$_{20}$ is already quite high within
the 3 ps of their simulation, but such a short simulation time led to a
value of $\delta \approx 0.1$.
If we average over the first 3 ps of our run performed at
T=324 K, we obtain $\delta$=0.15, in contrast with the converged value
$\delta$=0.29. 
A simulation time of 3 ps, corresponding to $\sim$ 5 vibrational periods, does
not allow for enough jumps to obtain a converged value of $\delta$.

The melting-like transitions in both Na$_8$ and Na$_{20}$ are found to be
spread over a broad temperature range, and in the case of Na$_{20}$ we have
identified a stepwise melting mechanism. Recently, Rey {\em et al.} \cite{Rey98}
in a study of the melting-like transition of a 13 particle cluster
associated the occurrence of melting in steps to the softness of the
repulsive core interaction rather than any many-body character of the
interactions. For the softer potentials considered,
two abrupt increases of $\delta$ were found: the first corresponding to the
onset of isomerization transitions involving only surface atoms, and the
second to complete melting. The stepwise melting mechanism was less clearly
defined for the harder interatomic potentials.
We have used our approximate functional in a series of static
calculations for Na$_2$ at several interatomic distances in order to construct
the binding energy curve, and have compared it to the interatomic potentials
considered by Rey {\em et al.} \cite{Rey98} We find that the repulsive
part of the binding energy curve obtained using the GEPS is softer than
their softest interatomic potential, and so the stepwise melting we have 
found is consistent with their findings.

Jellinek {\em et al.} \cite{Jel98} have demonstrated for Li$_8$ that the 
temperature at which exploration of low-lying isomers
begins is estimated more accurately in MC than in MD simulations. This is
because the J-walk MC sampling of the configurational space
is not hindered by the presence of energy barriers.
If the energy in a microcanonical MD simulation is just slightly higher
than the energy needed to surmount a barrier, exceedingly long simulation times
will be needed in order to achieve an efficient sampling. Thus, the real
temperatures at which the onset of melting occurs may be slightly
lower than those determined in our ``short'' simulations. The location of the
specific heat peaks, however, should be more accurate, because these occur for
a temperature higher than that corresponding to the onset of melting, where
the energy is already large enough to easily surmount the barriers. In summary,
the {\em width} of the specific heat peaks may be underestimated in MD
simulations.

We have studied the melting of clusters upon heating, but some differences
might be observed if the clusters were cooled from a high temperature
simulation. The details of the melting transition, such as the number and 
location of the steps in the $\delta$ curve, depend on the specific isomer from
which the dynamics is started, and thus may differ from those obtained 
through a cooling procedure. This is worthy of further study.

\section{Summary}

We have applied an {\em ab initio}, density-based, molecular
dynamics method to the study of melting
transitions in finite atomic systems. The computational effort which is
required is modest in comparison with that required by the traditional
CP-MD technique based on the use of the KS one-particle orbitals.
The computational effort to update the electronic system
scales linearly with the system size N, in contrast to the
N$^3$ scaling of orbital-based methods. This saving allows the method to be
used in the study of large clusters, and also for performing
extensive MD simulations in much the same way as the usual procedures 
involving phenomenological potentials.

The method has been applied here to study the melting-like transitions in
Na$_8$ and Na$_{20}$ clusters. Na$_8$ melts in a broad temperature
range from 110--220 K. The transition at $\sim$ 100 K is from a rigid
cluster in which the atoms are vibrating around their fixed equilibrium
positions to a phase in which the different permutational isomers of the GS
structure are visited. For higher temperatures the cluster distorts more and
more from the GS structure and 
can be considered fully liquid at T$ \approx $220 K. 
Na$_{20}$ undergoes several
successive transformations with increasing temperature, and the melting
transition is also spread over a wide temperature range.
In the first transition at T$_{m_1} \approx 110$ K surface atoms swap
positions without destroying the underlying icosahedral symmetry.
At T$_{m_2} \approx 160$ K, one of the inner atoms becomes surface-like,
the other remains in its central position for most of the simulation time,
and the surface of the cluster has melted. Swapping between the central
atom and one of the surface atoms is more frequent upon heating, until at
T$_{m_3} \approx 220$ K the value of $\delta$ saturates.
As expected, the temperatures of the melting intervals are lower than
the experimental melting temperature of bulk Na (T$_m^{bulk}$ = 371 K).
Preliminary calculations to simulate the melting of bulk Na
give a melting temperature in very good agreement with experiment, which
gives us confidence in the quantitative results we obtained for the
melting temperatures of the clusters.

Our findings suggest that it is important to study several melting
indicators when disentangling the details of melting-like transitions.
If a stepwise mechanism is present, the location of the
different steps are clearly indicated in $\delta$. The variation of
the diffusion coefficient with temperature obtained from the R(t) curves
is also useful in this respect. The ``atomic equivalence indices'' are
valuable for understanding the nature of
the structural transitions associated with each step in $\delta$.
We conclude that the specific heat alone may not be a sensitive indicator
of all the ``structural'' information contained in the $\delta$ curve.
Specifically, the final levelling off in $\delta$, marking the end of the
transition region and the temperature at which the liquid-like phase is
completely established, does not have any appreciable feature
associated with it in the specific heat. 

Comparison with other theoretical calculations performed at different
levels of theory suggests that our results agree with the {\em ab
initio} KS-LDA calculations of R\"othlisberger and Andreoni if the
same simulation times are considered. This agreement is 
strong validation of our model. However, we found it necessary to lengthen
the simulation times to at least 20 ps (and sometimes to 50--60 ps)
in order to obtain converged
global quantities. With the improved statistics, our results for Na$_8$
and Na$_{20}$ are in better agreement with those of Bulgac and coworkers and of
Calvo and Spiegelmann than
with those of Poteau {\em et al}.

$\;$

$\;$

$\;$

{\bf ACKNOWLEDGEMENTS:} This work has been supported by DGES 
(Grant PB95-0720-C02-01) , NATO(Grant CRG.961128) 
and Junta de Castilla y Le\'on (VA63/96 and VA72/96). We wish to thank
L. E. Gonz\'alez, M. J. L\'opez, and P. A. Marcos for useful discussions. A.
Aguado acknowledges a graduate fellowship from Junta de Castilla y Le\'on. M. J. Stott
acknowledges the support of the NSERC of Canada.


{\bf Captions of Figures and Tables.}

{\bf Figure 1} Geometries of the ground state isomers of (a) Na$_8$ and (b)
Na$_{20}$; (c) is an example of
the (19+1) structure that appears after the second
phase transition in Na$_{20}$ (see text for details).

{\bf Figure 2} (a) Specific heat and (b) relative rms bond
length fluctuation of Na$_8$ 
as functions of the internal temperature. The deviation around the mean
temperature is smaller than the size of the circles.
(c) Mean square displacement as a function
of time for three representative values of T.

{\bf Figure 3} Short-time averaged atomic equivalence indices
as functions of time for Na$_8$ at T= 34 K (a),
111 K (b), 220 K (c). Note the different simulation times for different 
temperatures.

{\bf Figure 4} (a) Specific heat and (b) relative rms bond
length fluctuation of Na$_{20}$  as functions of the internal temperature.
The deviation around the mean temperature is smaller 
than the size of the circles.
(c) Mean square displacement as a function
of time for four representative values of T.

{\bf Figure 5} Partial mean square displacements for core
(R$_c$(t)) and surface (R$_s$(t)) atoms in
Na$_{20}$, for T=109 K. 

{\bf Figure 6} Short-time averaged atomic equivalence indices
as functions of time for Na$_{20}$ at T= 32 K (a),
137 K (b),156 K (c). Note the different simulation times for each 
temperature.

{\bf Table 1} Bond length and binding energy per atom for Na$_2$ and
Na$_6$, calculated by Perdew and coworkers \cite{Nog96}  using Kohn-Sham
all electron (KSAE), and Kohn-Sham pseudopotential (KSPS) methods,
compared to our results of the gradient expansion functional of eq.
(13) with the same pseudopotential (GEPS). 



\begin{table}
\begin{center}
\begin{tabular}{c c c c c}
 &
\multicolumn {2}{c}{Bond Length (a.u.)} &
\multicolumn {2}{c}{Binding Energy (eV)} \\
\hline

 & Na$_2$ & Na$_6$ & Na$_2$ & Na$_6$ \\
\hline

KSAE & 5.64 & 6.51 & 0.44 & 0.63 \\
KSPS & 5.77 & 6.87 & 0.46 & 0.53 \\
GEPS & 5.99 & 6.81 & 0.55 & 0.69 \\
\hline

\end{tabular}
\end{center}
\caption{}
\end{table}

\newpage

\onecolumn[\hsize\textwidth\columnwidth\hsize\csname
@onecolumnfalse\endcsname

\begin{figure}
\psfig{figure=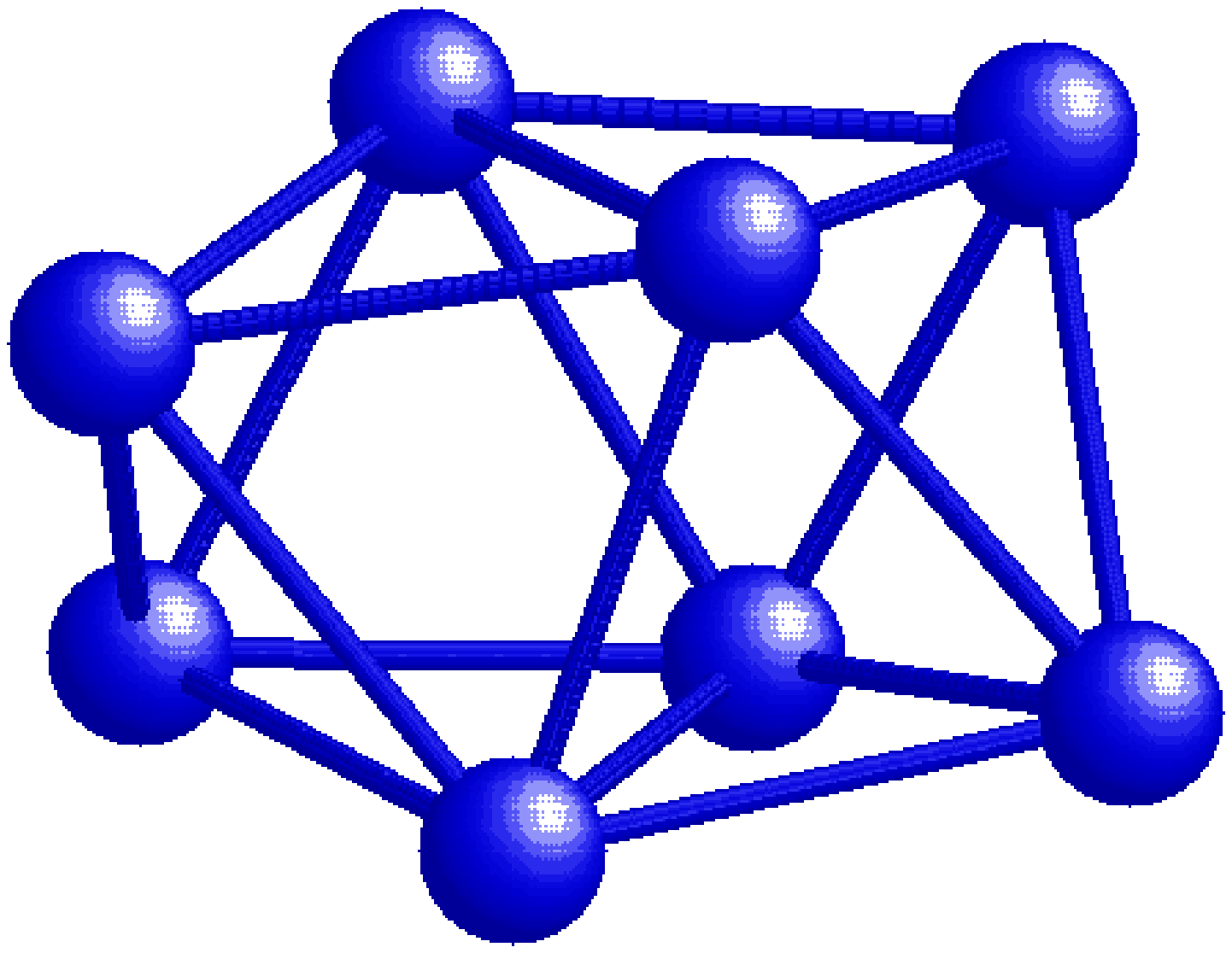}
\end{figure}

\begin{figure}
\psfig{figure=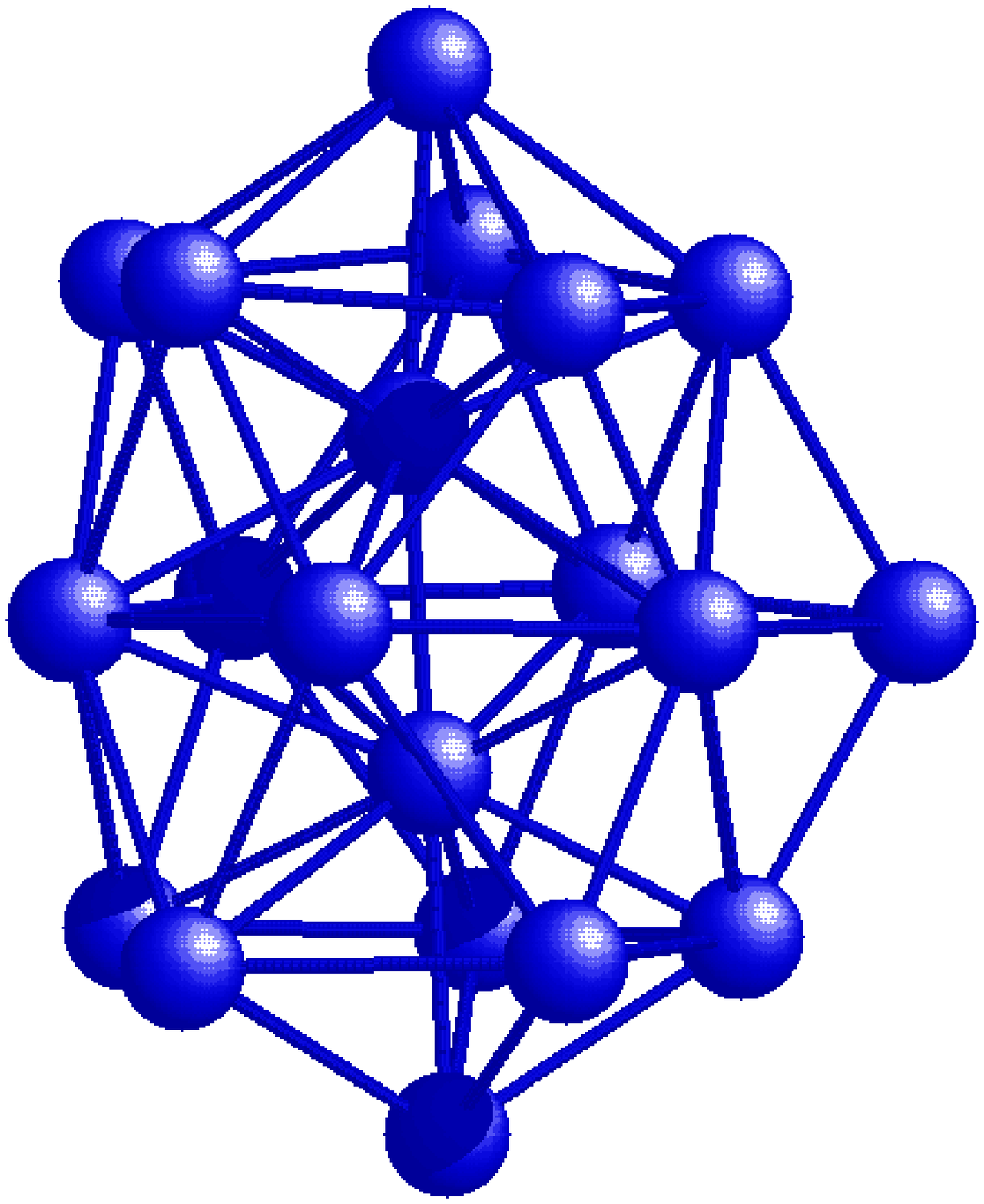}
\end{figure}

\begin{figure}
\psfig{figure=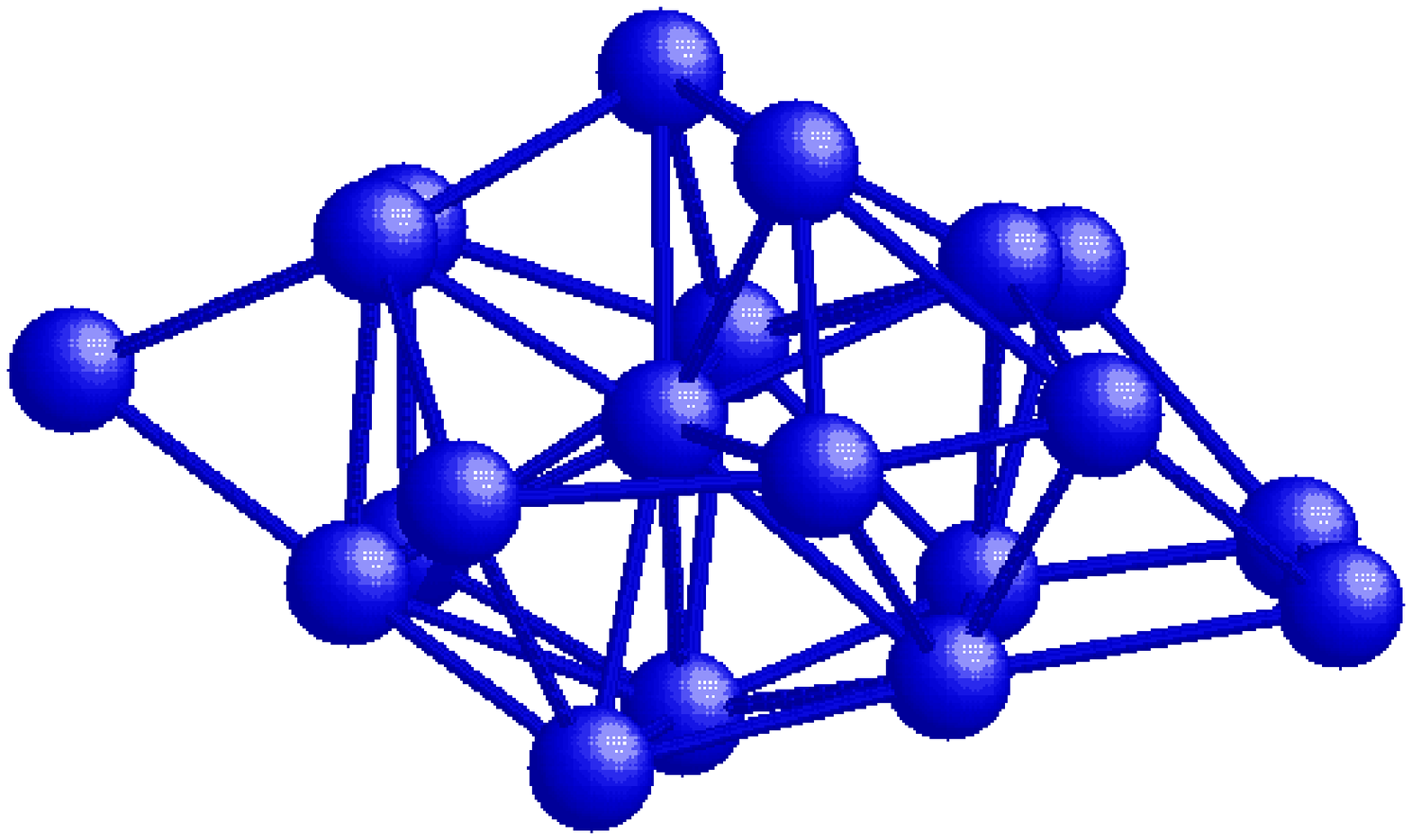}
\end{figure}

\begin{figure}
\psfig{figure=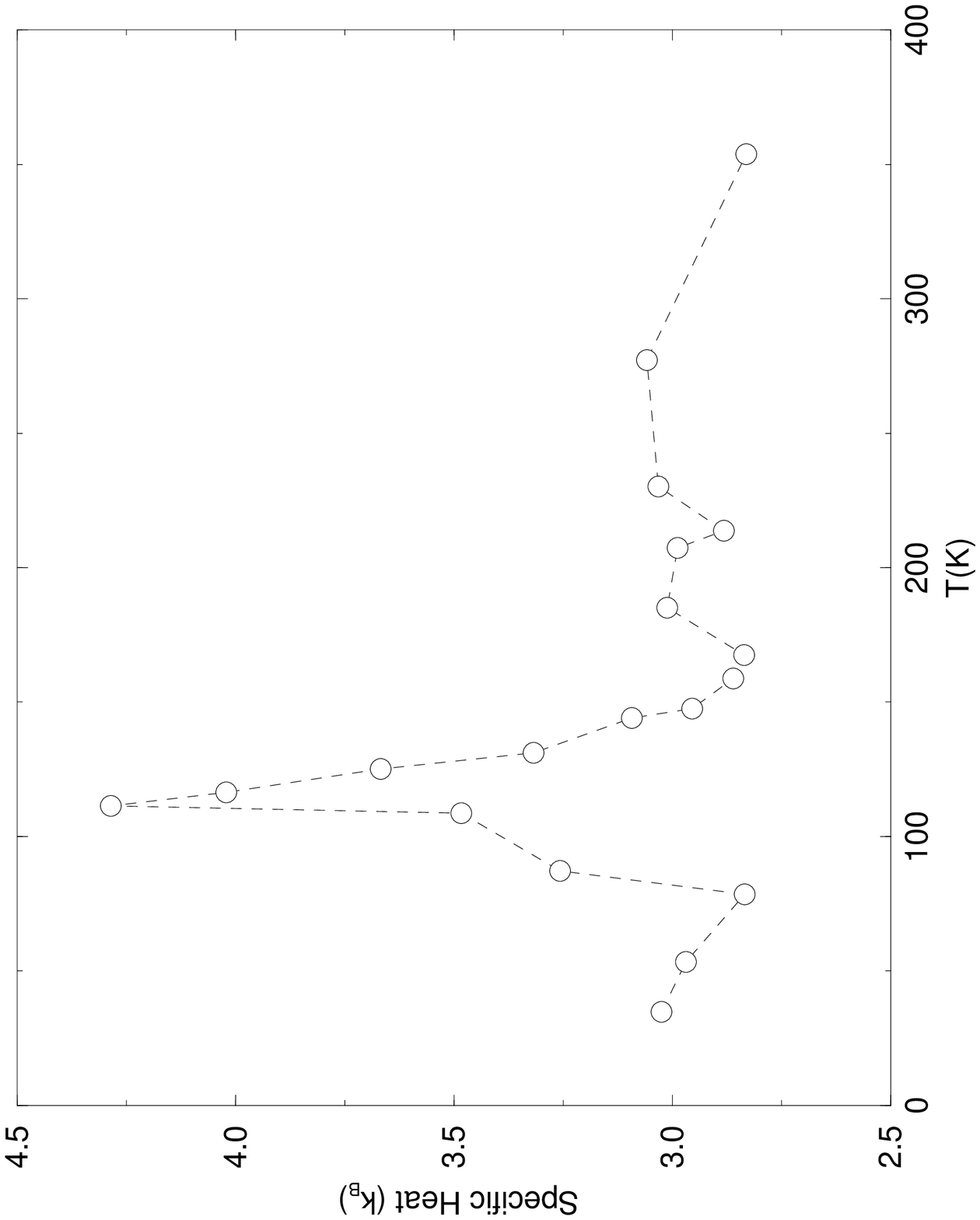}
\end{figure}

\begin{figure}
\psfig{figure=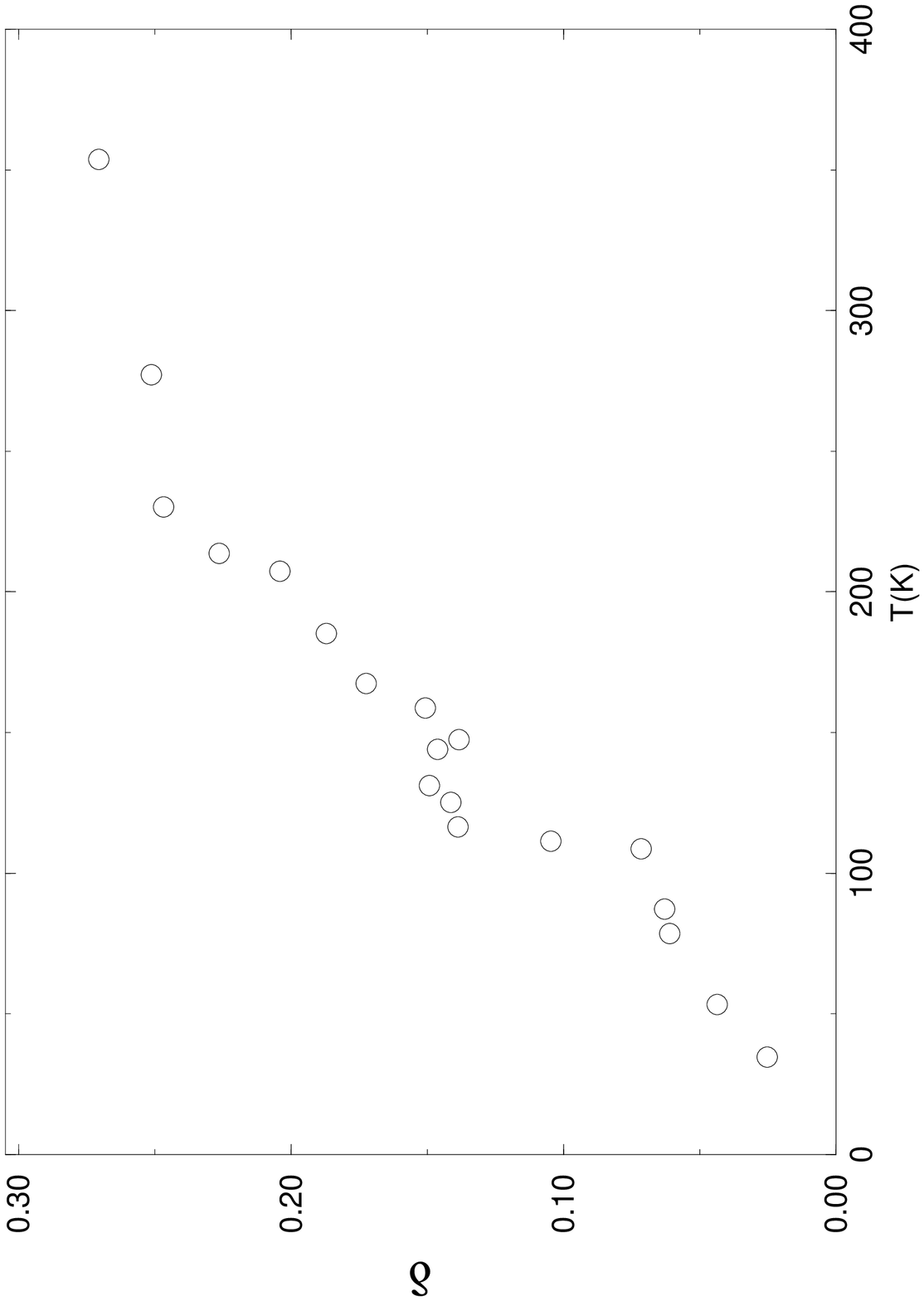}
\end{figure}

\begin{figure}
\psfig{figure=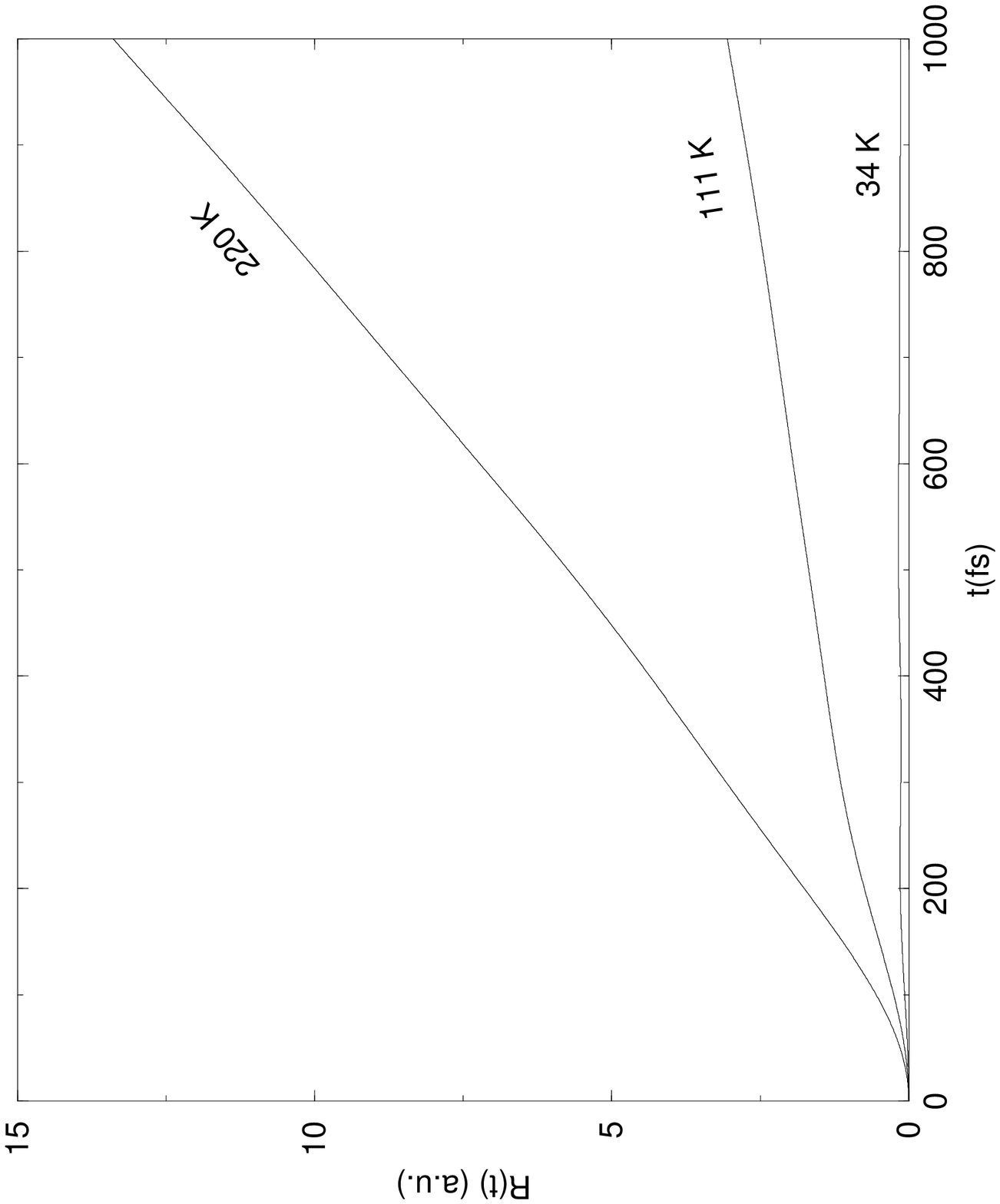}
\end{figure}

\begin{figure}
\psfig{figure=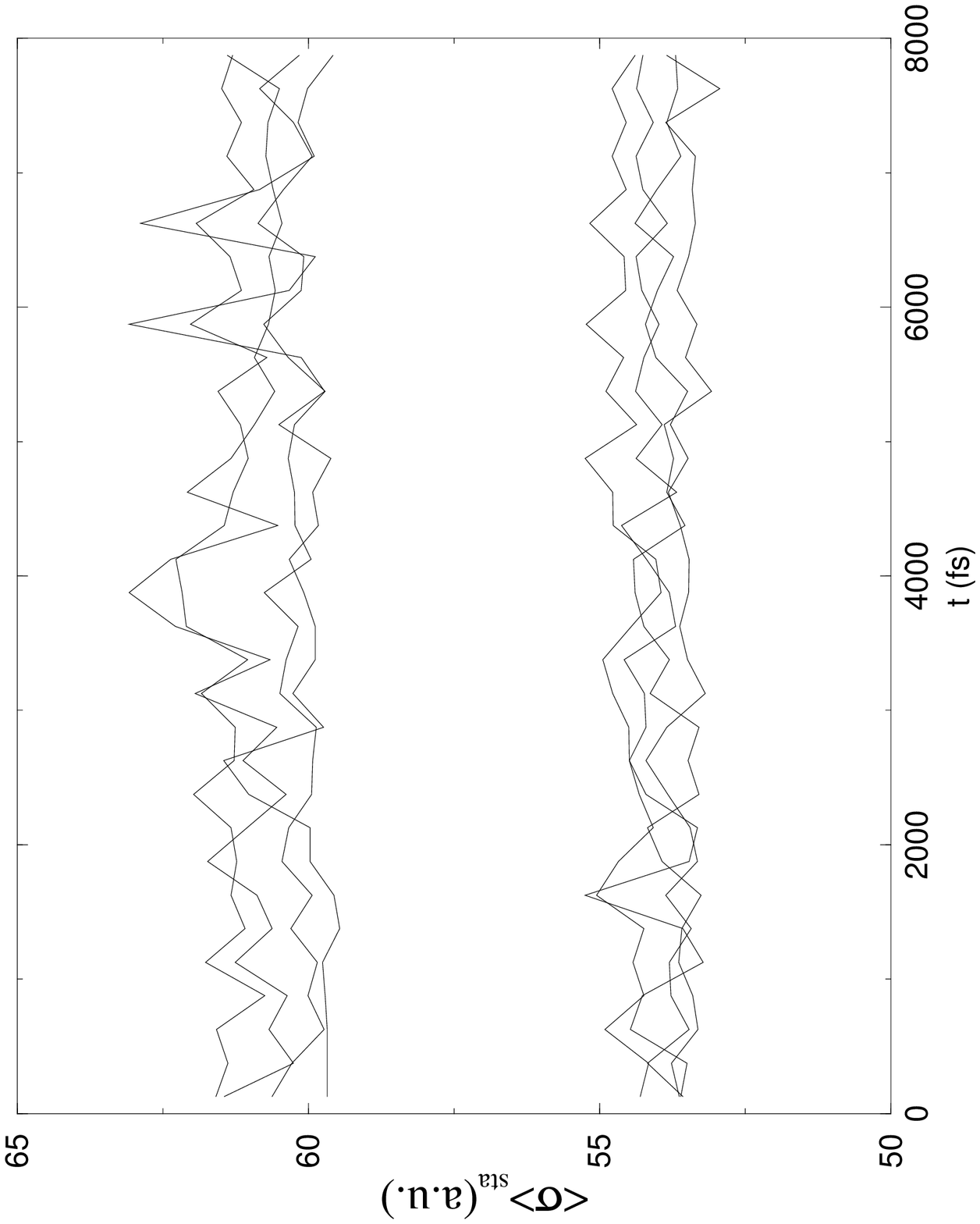}
\end{figure}

\begin{figure}
\vspace{-10mm}
\psfig{figure=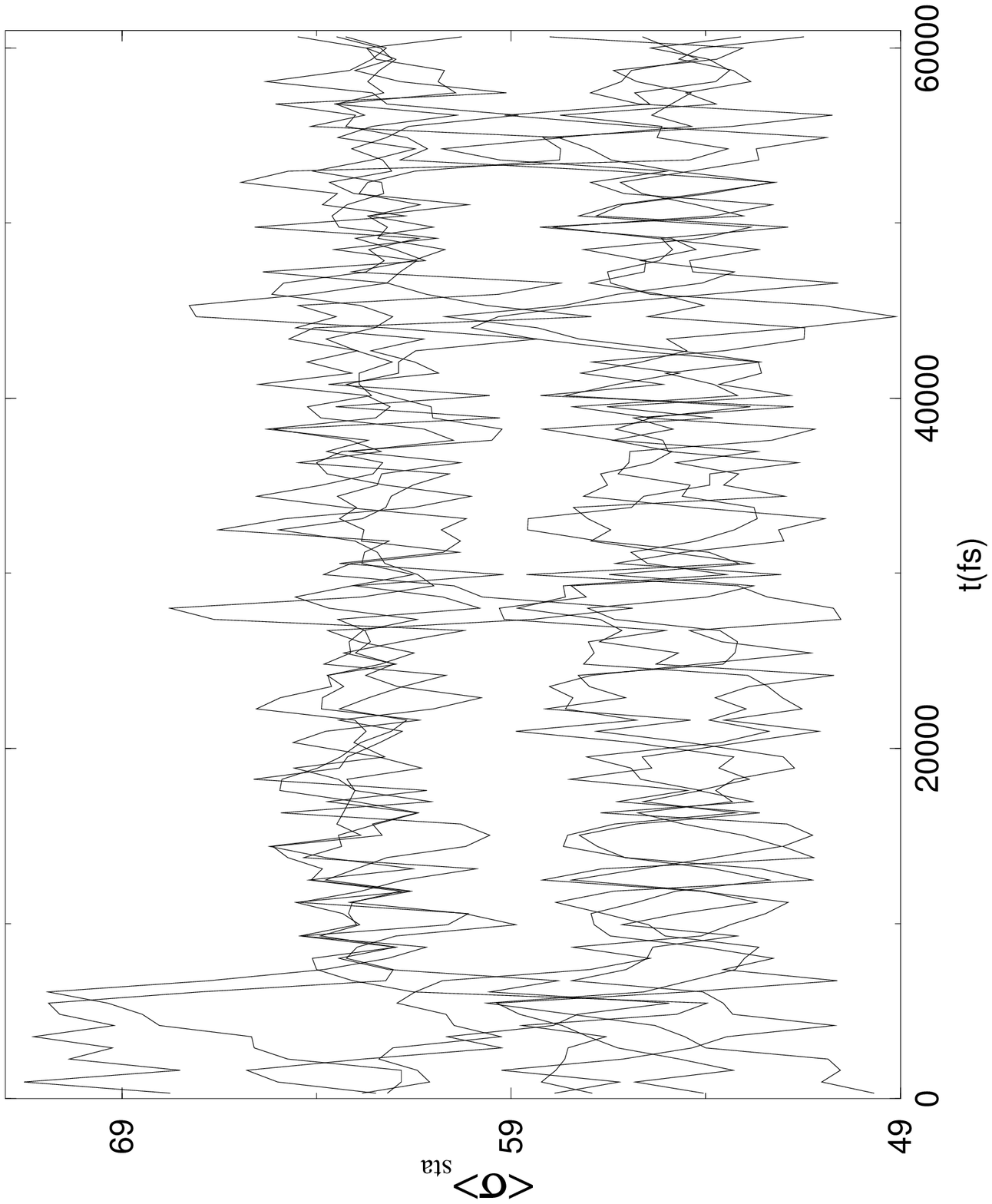}
\end{figure}

\begin{figure}
\psfig{figure=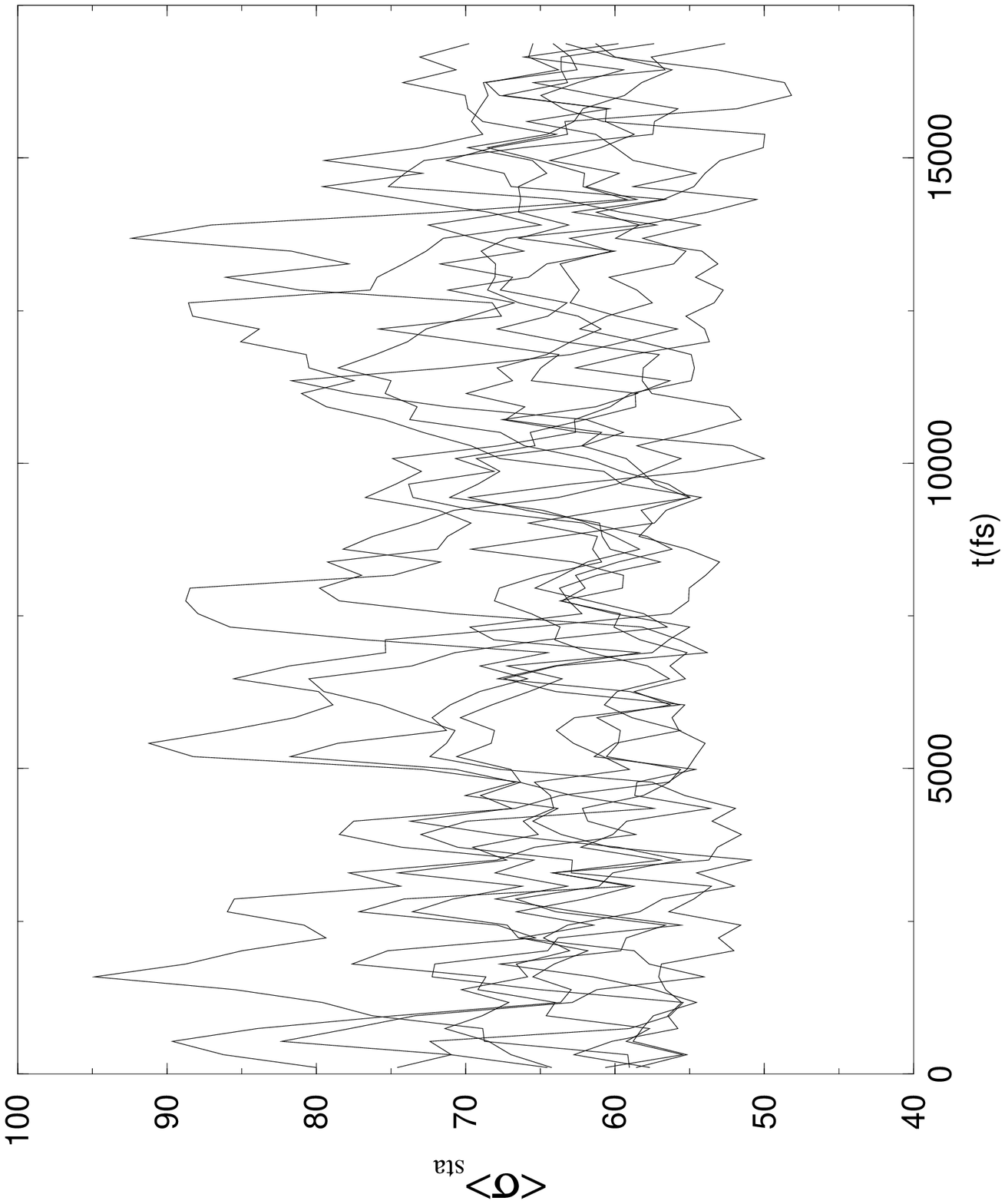}
\end{figure}

\begin{figure}
\psfig{figure=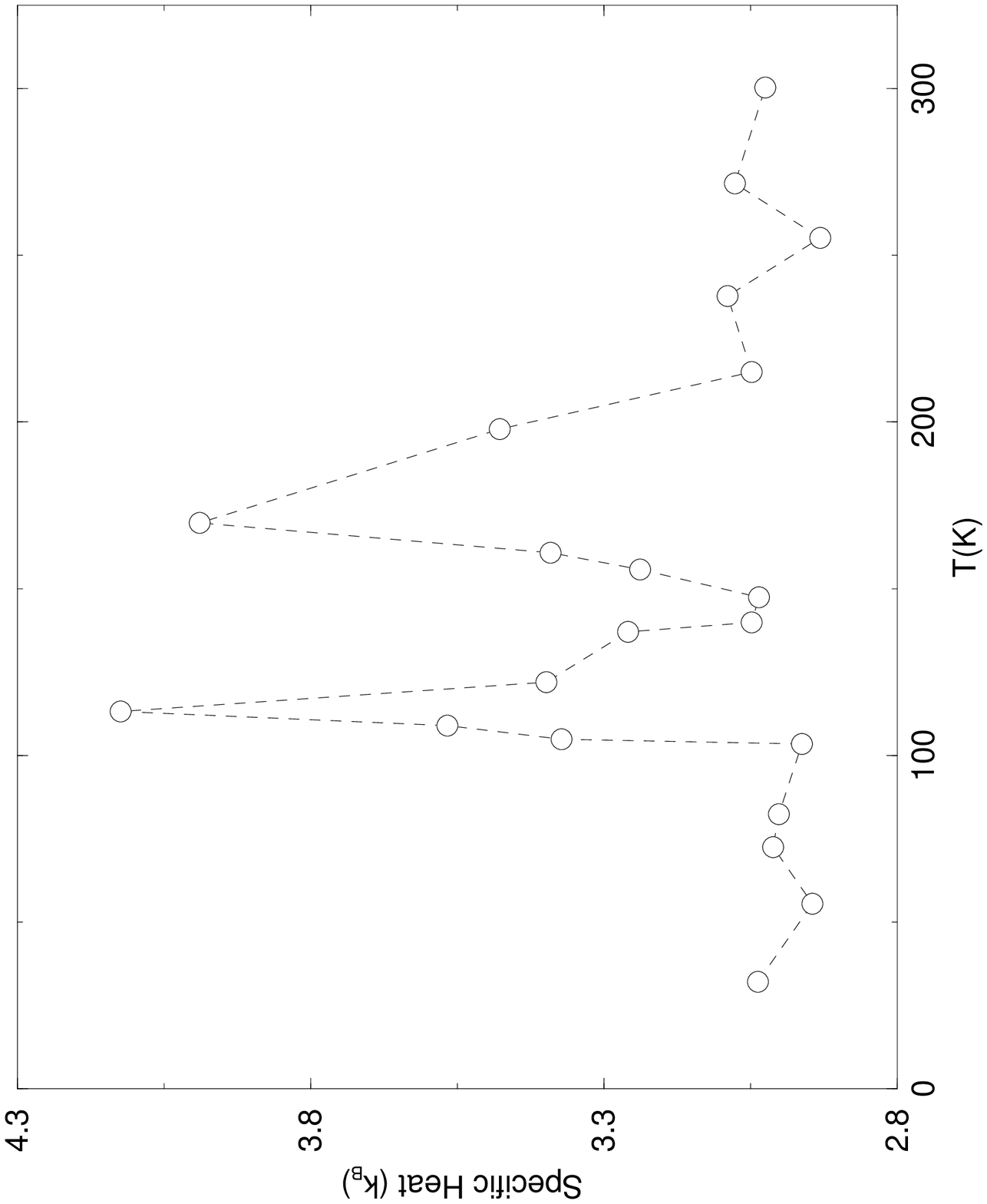}
\end{figure}

\begin{figure}
\psfig{figure=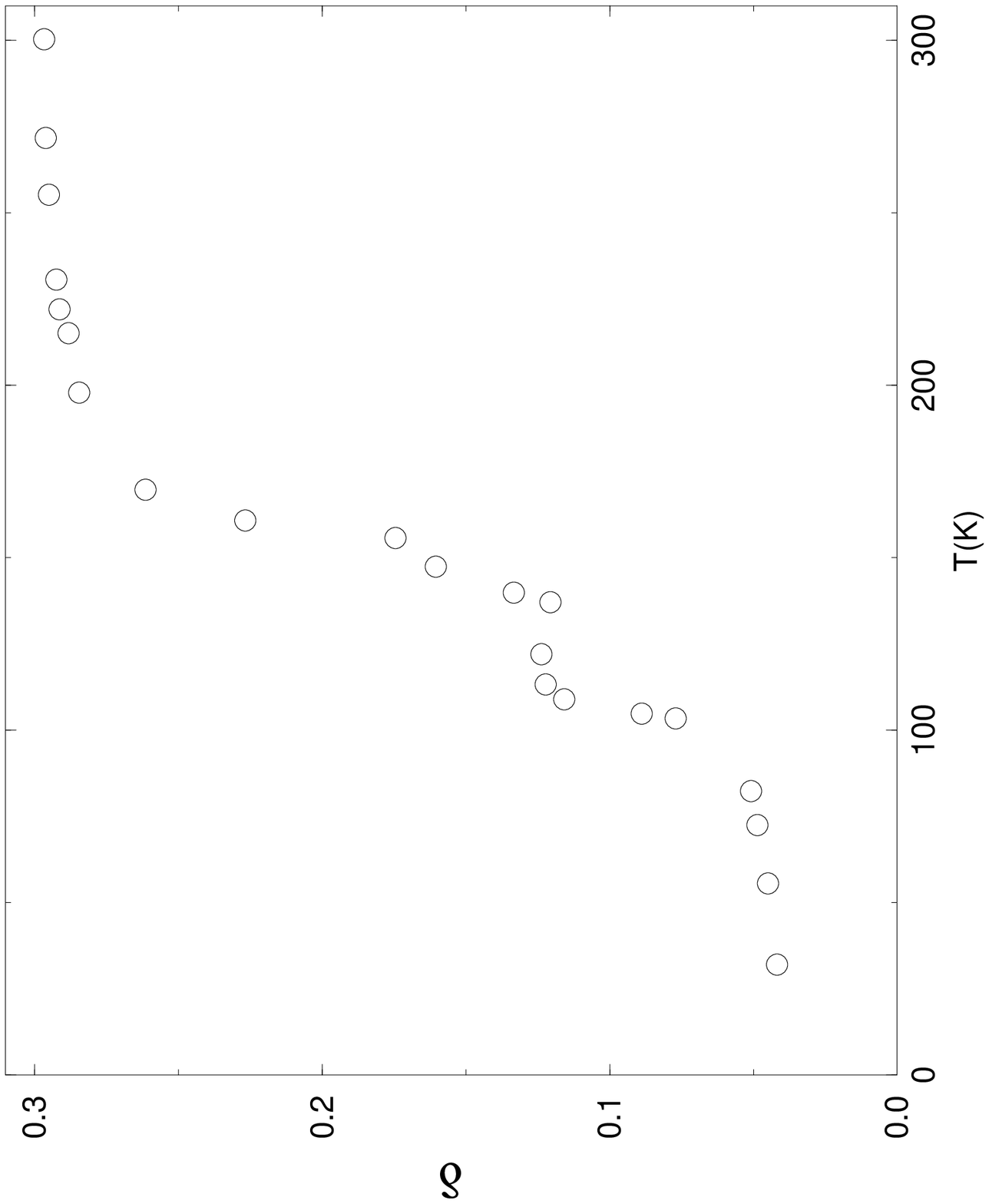}
\end{figure}

\begin{figure}
\psfig{figure=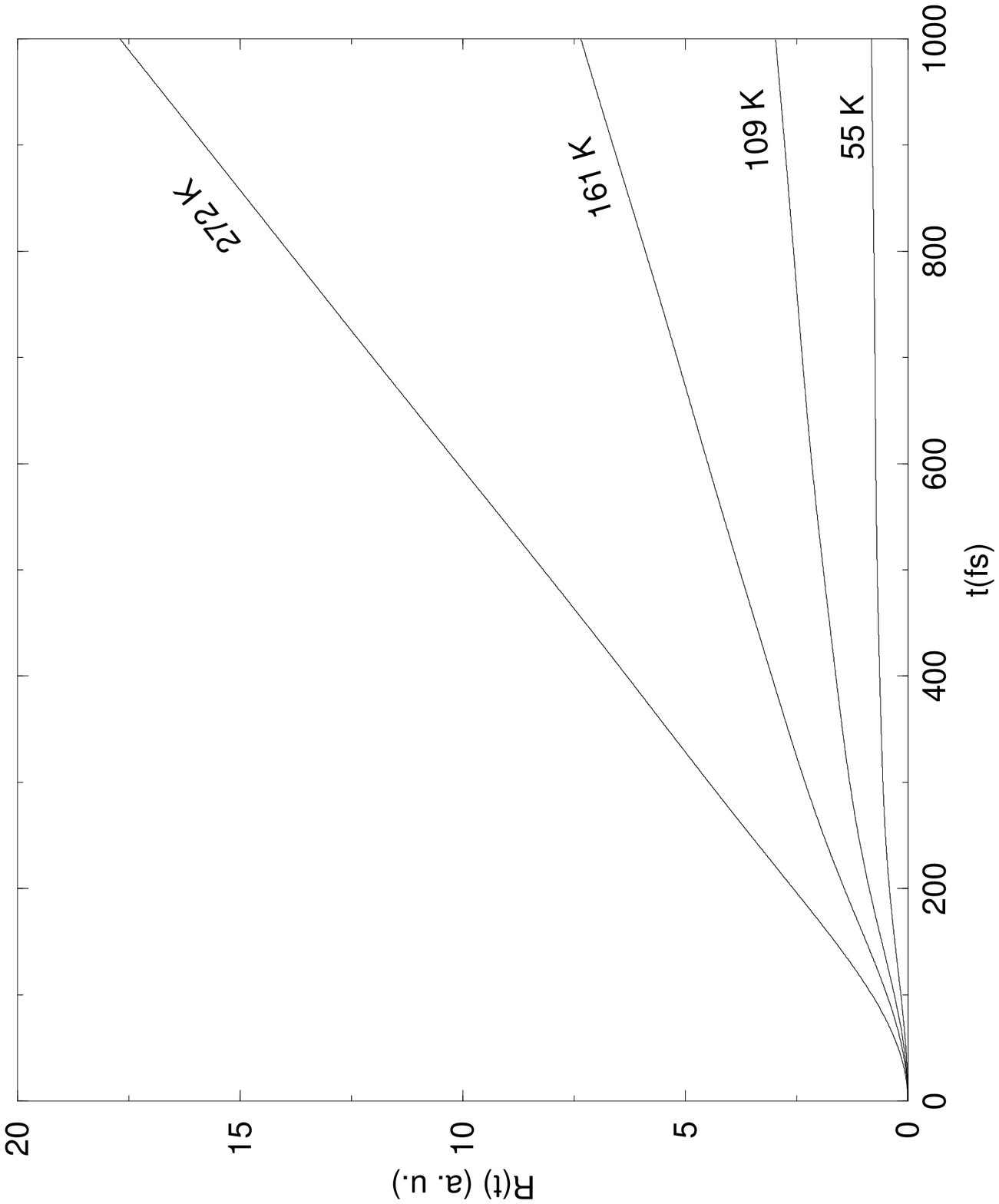}
\end{figure}

\begin{figure}
\psfig{figure=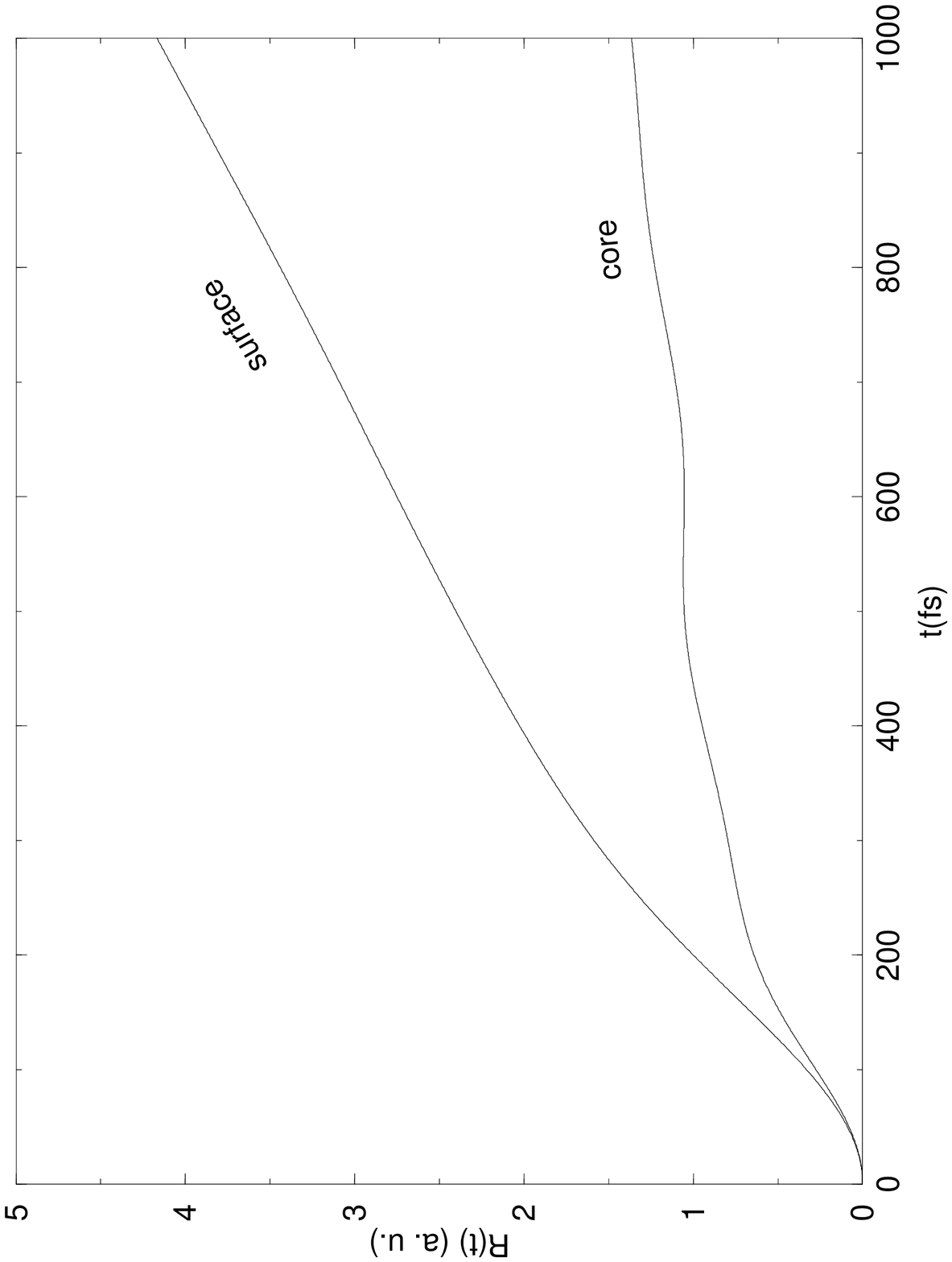}
\end{figure}

\begin{figure}
\psfig{figure=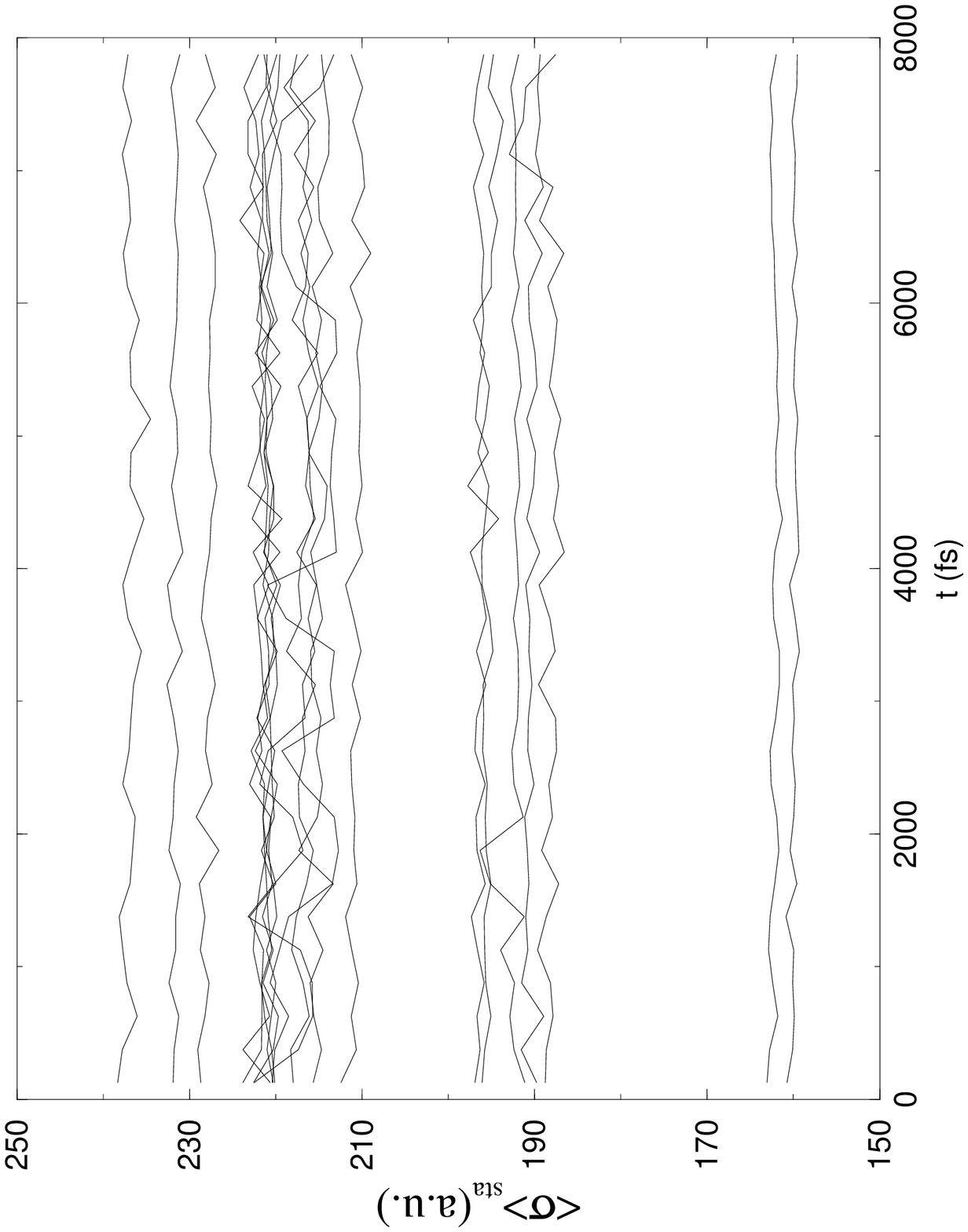}
\end{figure}

\begin{figure}
\psfig{figure=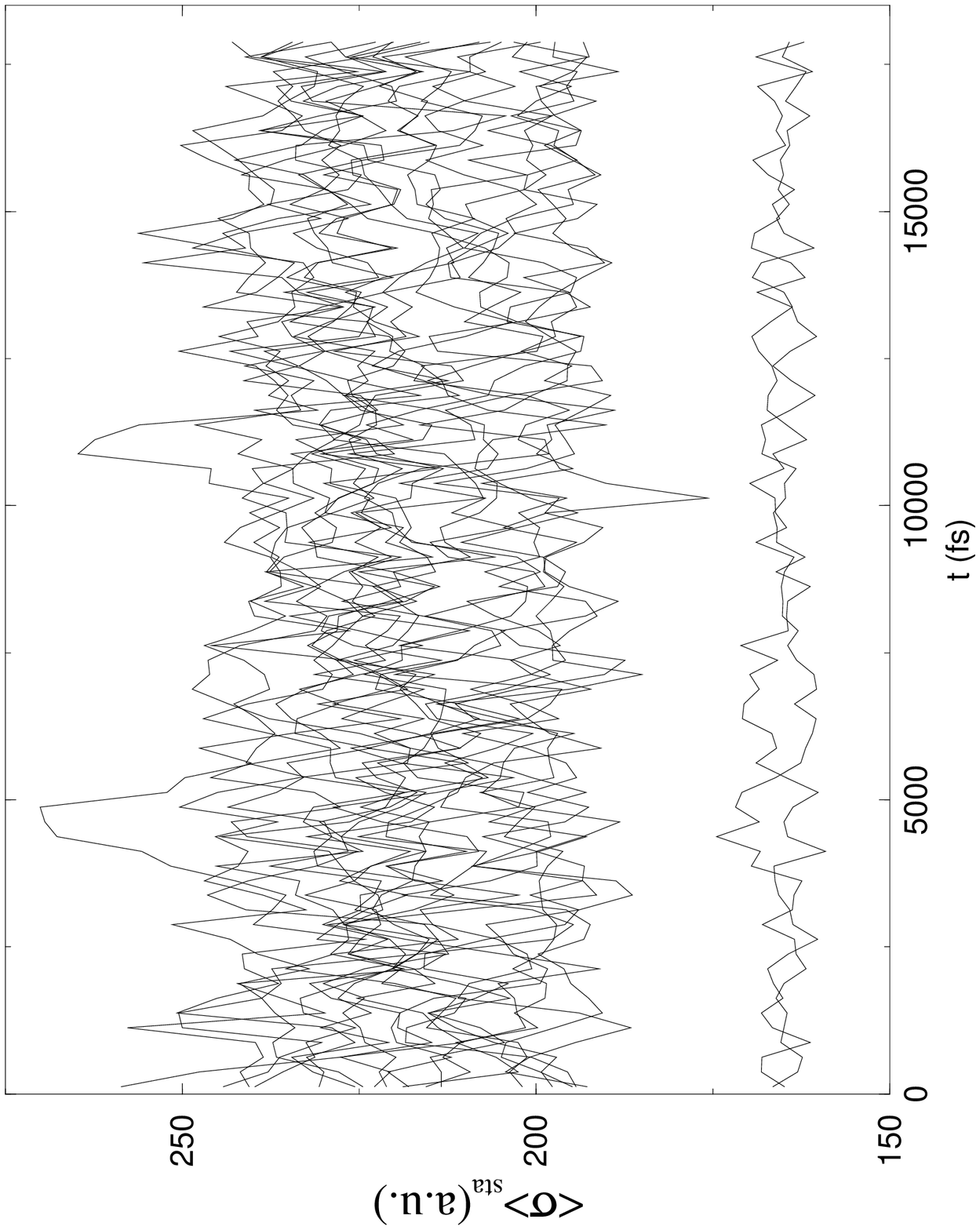}
\end{figure}

\begin{figure}
\psfig{figure=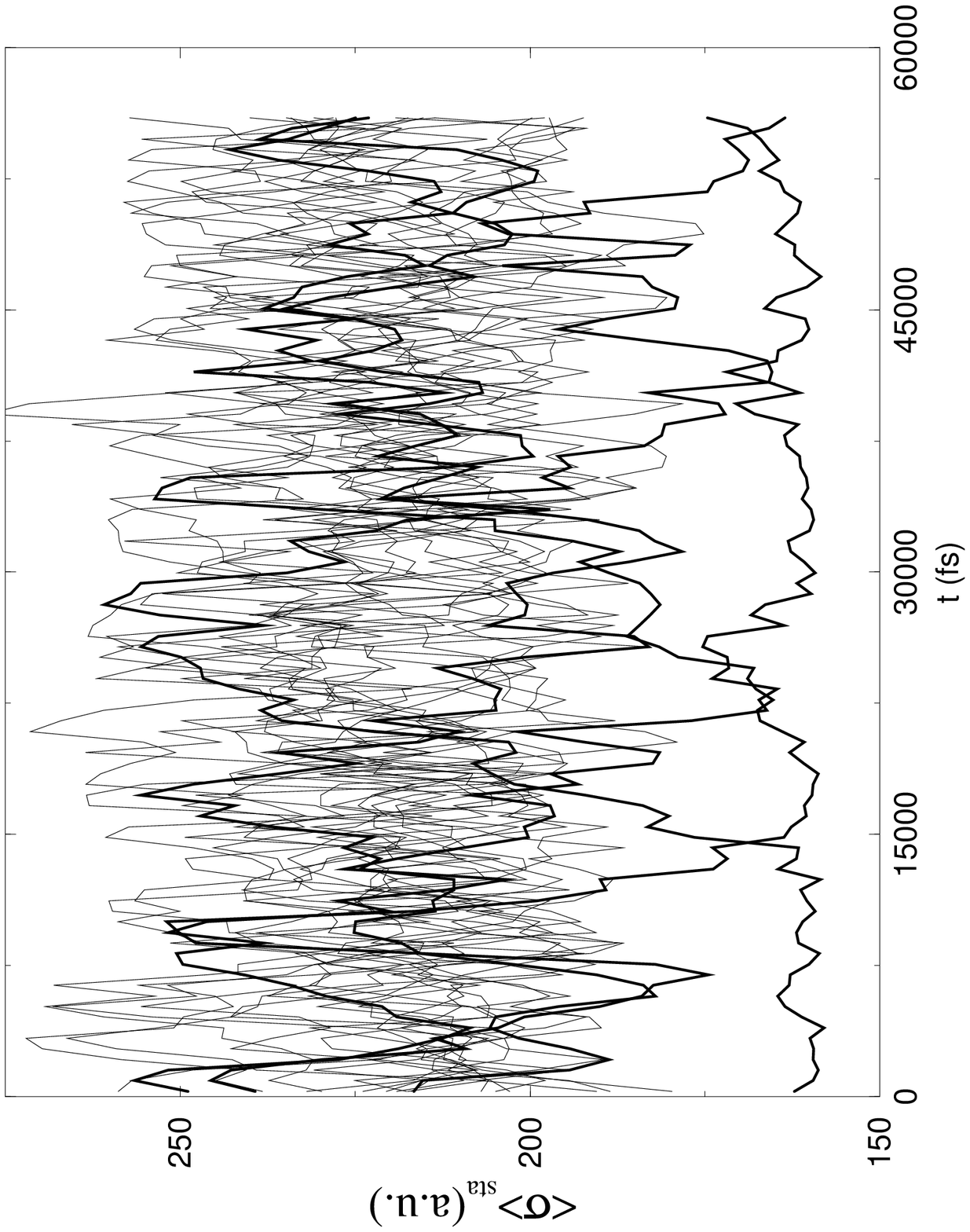}
\end{figure}

\end{document}